\documentclass{elsart}
\usepackage{graphicx}
\usepackage{literat}
\usepackage{amsmath}
\usepackage{amssymb}
\usepackage{epsfig}
\usepackage{dcolumn}
\usepackage{bm}
\newcommand{\D}{{\rm d }}
\newcommand{\ve}{{\varepsilon }}
\begin{document}
\begin{frontmatter}

\title{Entropy-driven phase transitions with influence of the field-dependent diffusion coefficient}

\author{V.O.~Kharchenko}
\ead{vasiliy@imag.kiev.ua}
\address{Institute of Magnetism, Nat. Acad. Sci. of Ukraine \break
36-B, Vernadsky St., 03680 Kyiv, Ukraine}

\date{\today}

\begin{abstract}

We present a comprehensive study of the phase transitions in the single-field
reaction-diffusion stochastic systems with field-dependent mobility of a
power-low form and the internal fluctuations. Using variational principles and
mean-field theory it was shown that the noise can sustain spatial patterns and
leads to disordering phase transitions. We have shown that the phase
transitions can be of critical or non-critical character.

\end{abstract}


\end{frontmatter}

\section{Introduction}

The ability of noise to induce spatial ordering has received a special
attention in last two decades. It is well known that in many situations noise
can actually play a constructive role, to name just a few one can consider:
noise-induced transitions in zero-dimensional systems \cite{Horstshemke}, noise
induced phase transitions in extended systems \cite{4NFPE3}, patterns formation
\cite{SSGO2007}, coupled Brownian motors \cite{11PRE2005,12PRE2005}, etc. Most
of the noise-induced effects are caused by an external fluctuations and
attributed to a short time instability. In particular, noise-induced patterns
and the phase transitions usually have a dynamical origin. It is principally
important that in the case of the internal noise, obeying
fluctuation-dissipation relation, a spatio-temporal coherence in nonlinear
systems can not be observed in a short-time scale. Such ordering processes
follow an entropy-driven mechanism when a kinetic coefficient/mobility is a
function of a stochastic field. It leads to the fact that a stationary
distribution is described by a free energy functional reduced to a Lyapunov
functional for deterministic dynamics and by an entropy contribution related to
the field-dependent mobility \cite{GO2001,katya,PhysRev2005,PRE2006}. It was
shown by the most of works concerning such phenomena that quantitative change
of the system behaviour is caused by the entropy variability.

It was shown that internal fluctuations can sustain the spatial coherence
(patterns) in nonlinear reaction-diffusion stochastic systems
\cite{PhysRev2005,PRL2006}. In the simplest case such models have two essential
features: the local dynamics and a transport phenomena. First of them is
determined by the chemical reactions in the system and the second one relates
to a diffusion of chemical species. Introducing a fluctuating source into the
model, we arrive at the stochastic description of non-equilibrium system
behaviour with possible spatial order. In this work we consider a prototype
model of the reaction-diffusion systems where an internal fluctuations are of a
multiplicative character (the noise intensity depends on a concentration field)
(see Ref.\cite{PhysRev2005} and citations therein). In the most of works
devoting to study the system behaviour with a field-dependent mobility, the
corresponding functional dependence is as follows: in the mixed state the
kinetic coefficient is large, whereas in dense ones it has small values. A
deterministic analysis of the systems with a generalized field-dependent
mobility was performed in Ref.\cite{Bray}, where the form of the kinetic
coefficient was proposed by authors solely. In our consideration we use a
generalized approach basing on a nonlinear master equation proposed by
G.~Kaniadakis \cite{Kaniadakis2001} and derive the generalized form for the
kinetic coefficient. Introducing the corresponding Fickian diffusion term into
the model and a term to describe a local dynamics of the field variables, we
arrive at a deterministic model for the reaction-diffusion systems. Considering
a system in real conditions we take into account a fluctuation source related
to an obtained mobility.

According to the well known approaches of entropy-driven phase transitions
\cite{GO2001,katya} one can investigate a possibility of patterns formation and
corresponding phase transitions. Our goal in this article is to perform a
detailed study of spatial system's coherent behaviour with a generalized
kinetic coefficient in a bistable stochastic system. With help of variational
principles we investigate a possibility of noise to sustain stationary spatial
structures. A picture of noise-induced phase transitions will be studied with
the mean-field approach and computer simulations.

Our paper is organized in the following manner. In the next section we present
the general formalism to investigate the noise-induced phase transitions and
patterns formation. In Section 3 we derive an expressions for the
field-dependent diffusion coefficient using the formalism of nonlinear master
equation and deformed Boltzmann-Gibbs statistics. According to the presented
formalism the noise-sustained structures are considered in Section 4. The
noise-induced phase transitions with the entropy-driven mechanism are studied
in Section 5. Main results and prospects for the future are presented in the
Conclusions.

\section{General theory}

Our starting point is the evolution equation for a particle density $p=p({\bf r
},t)$ written in a most general form of the continuity equation
\cite{Kaniadakis2001}
\begin{equation}
\frac{\partial p}{\partial t}=\nabla\left[D\gamma(p)\frac{\partial \ln
\kappa(p)}{\partial p}\nabla p\right]\label{NFPE0}.
\end{equation}
Here we take into account diffusion flux only with concentration-dependent
effective diffusion coefficient assumed in the form
\begin{equation}
D_{ef}(p)\equiv D\gamma(p)\frac{\partial \ln \kappa(p)}{\partial p},\quad
D={\rm const}. \label{D_eff}
\end{equation}
From a formal viewpoint the nonlinear diffusion equation (\ref{NFPE0}) can be
derived from a master equation assuming that probabilities of the microscopic
transitions explicitly depend on the concentrations of the initial and arrival
states, described by functions $\mu(p)$ and $\nu(p)$, respectively. It was
shown (see Refs.\cite{Kaniadakis2001Physa,Chavanis}) that functions $\gamma(p)$
and $\kappa(p)$ in Eq.(\ref{NFPE0}) are defined as follows:
$\gamma(p)=\mu(p)\nu(p)$, $\kappa(p)=\mu(p)/\nu(p)$. An explicit form for pairs
$(\mu,\nu)$ or $(\gamma,\kappa)$ can be set due to the physics of the problem
under consideration.

Introducing chemical reactions described by a rate $f(p)$, a generalized
continuity equation for the concentration of particles reads:
\begin{equation}
\frac{\partial p}{\partial t}=f(p)+\nabla\left[D_{ef}(p) \nabla
p\right]\label{NFPE1}.
\end{equation}
Formally, Eq.(\ref{NFPE1}) can be written with help of the Lyapunov functional
for the deterministic dynamics. Indeed, introducing notation for such a
functional in the form
\begin{equation}
F=\int\D {\bf r}\left\{\frac{1}{2}\left[D_{ef}(p)\nabla
p\right]^2+\varphi(p)\right\}
\end{equation}
with
\begin{equation}
\varphi=-\int\limits_0^pf(p')D_{ef}(p')\D p',
\end{equation}
the deterministic evolution equation (\ref{NFPE1}) takes a variational form
\cite{PhysRev2005}
\begin{equation}
\frac{\partial p}{\partial t}=-\frac{1}{D_{ef}(p)}\frac{\delta F[p]}{\delta p}.
\end{equation}

Considering the system in realistic conditions one needs to introduce a
fluctuation source $\xi(p;{\bf r},t)$. In our stochastic analysis we assume
that such fluctuating source obeys the fluctuation-dissipation relation
\cite{fd_theorem} and has following properties:
\begin{equation}
\langle\xi(p;{\bf r},\ t)\rangle=0,\quad \langle\xi(p;{\bf r },\ t)\xi(p;{\bf
r'},\ t')\rangle=\frac{2\sigma^2}{D_{ef}(p)}\delta({\bf r}-{\bf
r'})\delta(t-t'),
\end{equation}
where $\sigma^2$ is an intensity of the corresponding internal multiplicative
noise. Formally, one can introduce an external noise related to fluctuations of
a control parameter addressed to a local dynamics. In this paper we study
influence of an internal fluctuations source only on the phase transitions
picture with the concentration-dependent diffusion coefficient and the chemical
reactions. In the following analysis we use the Stratonovich interpretation of
the Langevin equation
\begin{equation}
\frac{\partial p}{\partial t}=-\frac{1}{D_{ef}(p)}\frac{\delta F[p]}{\delta
p}+\frac{1}{\sqrt{D_{ef}(p)}}\tilde\xi({\bf r},\ t)\label{NFPE2},
\end{equation}
where $\xi(p;{\bf r},t)=[D_{ef}(p)]^{-1/2}\tilde\xi({\bf r},t)$.

Considering stationary properties of the system we exploit a stationary
probability density functional, obtained as a solution of the corresponding
Fokker-Planck equation \cite{Risken}. In the framework of standard technique
such stationary functional takes the form \cite{4NFPE3,GO2001,4NFPE1}
\begin{equation}
P_{st}\propto \exp\left(-U_{ef}[p]/\sigma^2\right)\label{Pst};
\end{equation}
the effective energy functional
\begin{equation}
U_{ef}[p]=F[p]-\frac{\sigma^2}{2}\int\D {\bf r}\ln D_{ef}(p)\label{U_eff}
\end{equation}
is defined through the free energy functional $F[p]$ and the effective
diffusion coefficient $D_{ef}(p)$. Due to the stationary probability density
functional (\ref{Pst}) has an exact form, the effective potential (\ref{U_eff})
can be used to study possibility of the structure formation under the
multiplicative noise influence. Moreover, we can apply the mean-field theory
formalism to investigate the noise-induced phase transitions in systems of such
a kind.

\section{Model}

The first problem we deal with is to set a generalized form for the effective
diffusion coefficient $D_{ef}(p)$, defined according to Eq.(\ref{D_eff}). The
main criterion for the function $D_{ef}(p)$ is a bell-shaped form in the
interval $p\in[0,1]$. We assume that following properties are satisfied:
$D_{ef}(0)=D_{ef}(1)=0$; $D_{ef}(1/2)=D_{ef}^{(max)}$, where $D_{ef}^{(max)}$
is a maximal value. It means that fluctuations are large in a mixed state
characterized by the value $p=1/2$; in dense states ($p=1$, $p=0$) no
fluctuations are realized. Considering a general problem, related to a complex
systems investigation, let us assume that the function $D_{ef}(p)$ has a
power-law form. Indeed, as it was shown before the complex systems can be
described by the nonlinear continuity equation and are characterized by a
power-law form for main statistical quantities (see
Refs.\cite{Plastino,Frank,LBorland,callis}). Most of complex statistical
systems are described by deformed Boltzmann-Gibbs statistics or q-statistics
exploiting deformed logarithm's and exponential functions \cite{callis}. By now
there are two well known kinds of deformations, proposing by G.Kaniadakis and
C.~Tsallis (see Refs.\cite{Kaniadakis2001Physa,callis}, respectively). In this
paper we operate with a mathematical construction of the q-deformed logarithm
$\ln x\to\ln_qx =(x^{1-q}-1)/(1-q)$ \cite{callis}, the exponent $q$ is a
non-additivity parameter; in the limit case $q=1$ one arrives at usual
logarithm and the standard Boltzmann-Gibbs statistics. Thus, inserting
q-deformed logarithm into Eq.(\ref{D_eff}), one has
\begin{equation}
D_{ef}(p)=D\gamma(p)\kappa(p)^{-q}\frac{\D\kappa(p)}{\D p}. \label{Def_gk}
\end{equation}
In the most of works related to the nonlinear continuity equation (\ref{NFPE0})
the functions $\gamma(p)$ and $\kappa(p)$ are assumed to be linear versus its
argument. In the simplest case one has: $\kappa(p)=p$, $\gamma(p)=p$. It
results to the construction of the form
\begin{equation}
D_{ef}(p)=Dp^{1-q}. \label{D_simplest}
\end{equation}
It is seen that the linear approximation of $\gamma(p)$ and $\kappa(p)$ does
not give the bell-shaped form for the effective diffusion coefficient.
Evolution of the system with the effective diffusion coefficient
(\ref{D_simplest}) was described in Ref.\cite{physa}. It was shown that such
complex systems manifest an anomalous diffusion.

To satisfy the main condition for the function $D_{ef}(p)$ let us assume
nonlinear constructions for $\gamma(p)$ and $\kappa(p)$. According to the fact
that complex systems are self-similar, let us write $\gamma(p)$ in the form
\begin{equation}
\gamma(p)=p^{\alpha}\label{gamma},
\end{equation}
where the exponent $\alpha\in(0,1)$. To define the function $\kappa(p)$ we
suppose that the derivative $\D\kappa(p)/\D p$ in Eq.(\ref{Def_gk}) gives a
power-low dependence, i.e. $\D\kappa(p)/\D p=\kappa^\beta$, $\beta>0$. Such an
assumption determines the power-low form for the function $\kappa(p)$. Indeed,
after some algebra one can obtain
$\kappa(p)=\left[C(\beta-1)+(1-\beta)p\right]^\frac{1}{1-\beta}$, where $C>0$
is a constant. Next, introducing a positive constant $C_1=C(\beta-1)$ with
$\beta>1$ one can represent the function $\kappa(p)$ in the form of the Tsallis
exponent \cite{Tsallis}:
\begin{equation}
\kappa(p)=[C_1+(1-\beta)p]^\frac{1}{1-\beta}\equiv
C_1^\frac{1}{1-\beta}\exp_{\beta}\frac{p}{C_1}.\label{kappa}
\end{equation}
The exponent $\beta$ plays a role of the non-additivity parameter. The Tsallis
exponent in Eq.(\ref{kappa}) becomes the usual one at $\beta=1$. Substituting
expressions for the functions $\gamma(p)$ and $\kappa(p)$ from Eq.(\ref{gamma})
and Eq.(\ref{kappa}), respectively, into the equation (\ref{Def_gk}) and
performing trivial calculations one arrives at the power-low form for the
effective diffusion coefficient:
\begin{equation}
D_{ef}(p)=Dp^\alpha\left[C_1-(\beta-1)p\right]^\delta,
\quad\delta=\frac{q-\beta}{\beta-1}.\label{Deff12}
\end{equation}
The function $D_{ef}(p)$ has a bell-shaped form only if $\delta>0$ or if
$q>\beta$ with $\beta>1$. Formally, the effective diffusion coefficient
$D_{ef}(p)$ defined by the exponent $\alpha\ne\delta$ is a non-symmetrical
function with respect to the point $p=1/2$. In further investigation we are
interested in studying the system properties when $D_{ef}(p)$ is the
symmetrical function. To this end we assume $\delta=\alpha$. It allows to
obtain the relation between the both non-additivity parameters $q$ and $\beta$
in the form
\begin{equation}
\beta<q<2\beta-1.
\end{equation}
Therefore, we arrive at the symmetrical form for the effective diffusion
coefficient
\begin{equation}
D_{ef}(p)=\frac{\beta_0}{2}p^\alpha\left[1-\frac{p}{p_s}\right]^\alpha,\label{Deff_final}
\end{equation}
where $p_s=C_1/(\beta-1)$ is a saturation concentration, $\beta_0=2DC_1^\alpha$
is a constant. The obtained formula (\ref{Deff_final}) can be derived directly
if probabilities $\mu(p)$ and $\nu(p)$ are known initially. Indeed, using the
approach developed in Ref.\cite{Chavanis} one can express functions $\gamma(p)$
and $\kappa(p)$ through $\mu(p)$ and $\nu(p)$, and after find the alternative
construction for the effective diffusion coefficient:
\begin{equation}
D_{ef}(p)=D\left(\frac{\mu(p)}{\nu(p)}\right)^{1-q}\left[\nu\frac{\D}{\D
p}\mu-\mu\frac{\D}{\D p}\nu\right].
\end{equation}
As it follows from our consideration densities $\mu(p)$ and
$\nu(p)$ can be obtained using relations between $\gamma(p)$ and
$\kappa(p)$. After a trivial algebra one find:
\begin{equation}
 \mu(p)=\left[\gamma(p)\kappa(p)\right]^{1/2},\quad
 \nu(p)=\left[\gamma(p)/\kappa(p)\right]^{1/2}.
\end{equation}
Therefore, the form for the effective diffusion coefficient is well defined.

Next, to derive a model for the function $f(p)$ that describes possible
chemical reactions in the system. Let us assume that in the deterministic
regime there are three stationary concentration values $p^{(1)}_0$,
$p^{(2)}_0$, $p^{(3)}_0$ with $p^{(i)}_0\ne p^{(j)}_0$, $i\ne j$. Thus, using
the theory of dynamical systems one can suppose the deterministic force in the
form $f(p)=-\prod_i(p-p^{(i)}_0)$. In the simplest case the following
construction can be used:
\begin{equation}
f(p)=\ve(p-\lambda)-(p-\lambda)^3,\label{f(p)}
\end{equation}
where $\ve$ is a control parameter, $\lambda$ determines the equilibrium
concentration magnitude. In such a case stationary states are determined by the
values: $p^{(1)}_0=\lambda$, $p^{(2,3)}_0=\lambda\pm\sqrt{\ve}$. In the
following analysis we choose $\lambda=1/2$.

Next, let us show that the corresponding internal multiplicative noise leads to
a short-time instability of the mixed/disordered state $p=1/2$. The linear
stability analysis can be performed for an auxiliary field $y({\bf r},t)=p({\bf
r},t)-\lambda$. The corresponding linearized Langevin equation takes the form
\begin{equation}
\frac{\partial y({\bf r},t)}{\partial t}=\ve
y+\varkappa^2\nabla^2y+\tilde{\sigma}^2y+\varkappa^{-1}\xi({\bf r },t),
\label{Langevin_linear}
\end{equation}
where $\varkappa^2=1/2\beta_0\lambda^{2\alpha}$,
$\tilde{\sigma}^2=(2\alpha\sigma^2)/(\beta_0\lambda^{2(\alpha+1)})$. In the
Fourier space Eq.(\ref{Langevin_linear}) is
\begin{equation}
\frac{\partial y(\pm {\bf k},t)}{\partial t}=\ve y(\pm {\bf
k},t)-\varkappa^2k^2y(\pm {\bf k},t)+\tilde\sigma^2y(\pm {\bf
k},t)+\varkappa^{-1}\xi(\pm {\bf k},t).
\end{equation}
Writing the dynamical equation for the two point correlation function
\begin{equation}
 \frac{\partial}{\partial t}\langle y({\bf k},t)y(-{\bf k},t)\rangle=
 \left\langle y({\bf k},t)\frac{\partial}{\partial t}y(-{\bf k},t)\right\rangle
 + \left\langle y(-{\bf k},t)\frac{\partial}{\partial t}y({\bf k},t)\right\rangle
\end{equation}
and calculating correlators $\langle \xi_{-{\bf k}}y_{{\bf k}}\rangle$,
$\langle \xi_{{\bf k}}y_{-{\bf k}}\rangle$ with the help of Novikov's theorem
\cite{Novikov13}, one arrives at a dynamical equation for the structure
function $S({\bf k},t)=\langle y_{{\bf k}}(t)\ y_{-{\bf k}}(t)\rangle$ in the
form
\begin{equation}
\frac{\partial}{\partial t}S({\bf k},t)=2(\ve - \varkappa^2{\bf k
}^2+\tilde\sigma^2)S({\bf k},t)+2\frac{\sigma^2}{\varkappa}.
\end{equation}
It is seen that the internal multiplicative noise leads to a short-time
instability of the mixed state $p=1/2$. The stationary value of the structure
function is
\begin{equation}
S({\bf k})=\frac{\sigma^2/\varkappa}{\varkappa^2 k^2-\ve-\tilde{\sigma}^2}.
\end{equation}
Hence, the homogeneous state $p({\bf r},t)=1/2$ is stable only if
$\ve<\varkappa^2k^2-\tilde\sigma^2$.

\section{Noise sustained structures}

As was shown in Ref.\cite{PhysRev2005} the extrema of the effective potential
$U_{ef}[p]$ correspond to the stationary noise-sustained structures $p_{st}$.
Such structures can be computed as an solution of equation $\delta
U_{ef}[p]/\delta p=0$, where the effective functional (\ref{U_eff}) has the
form
\begin{equation}
U_{ef}[p]= \int\D {\bf r}\left[\varphi(p)+\frac{1}{2}\left(D_{ef}(p)\nabla p
\right)^2\right]-\frac{\sigma^2}{2}\int\D {\bf r}\ln D_{ef}(p).
\end{equation}
Making the first variation of $U_{ef}[p]$ with respect to $p$ equal to zero, we
arrive at the equation for the stationary structures
\begin{equation}
\Delta p=\left.-\left[\frac{\sigma^2}{2D_{ef}^3(p)}\frac{\partial
D_{ef}(p)}{\partial p}+\frac{f(p)}{D_{ef}(p)}+\frac{1}{D_{ef}(p)}\frac{\partial
D_{ef}(p)}{\partial p }\left(\nabla p\right)^2\right]\right|_{p=p_{st}}.
\label{Delta_p}
\end{equation}

\begin{figure}[!ht]
\centering
 \small{a)}\includegraphics[width=65mm]{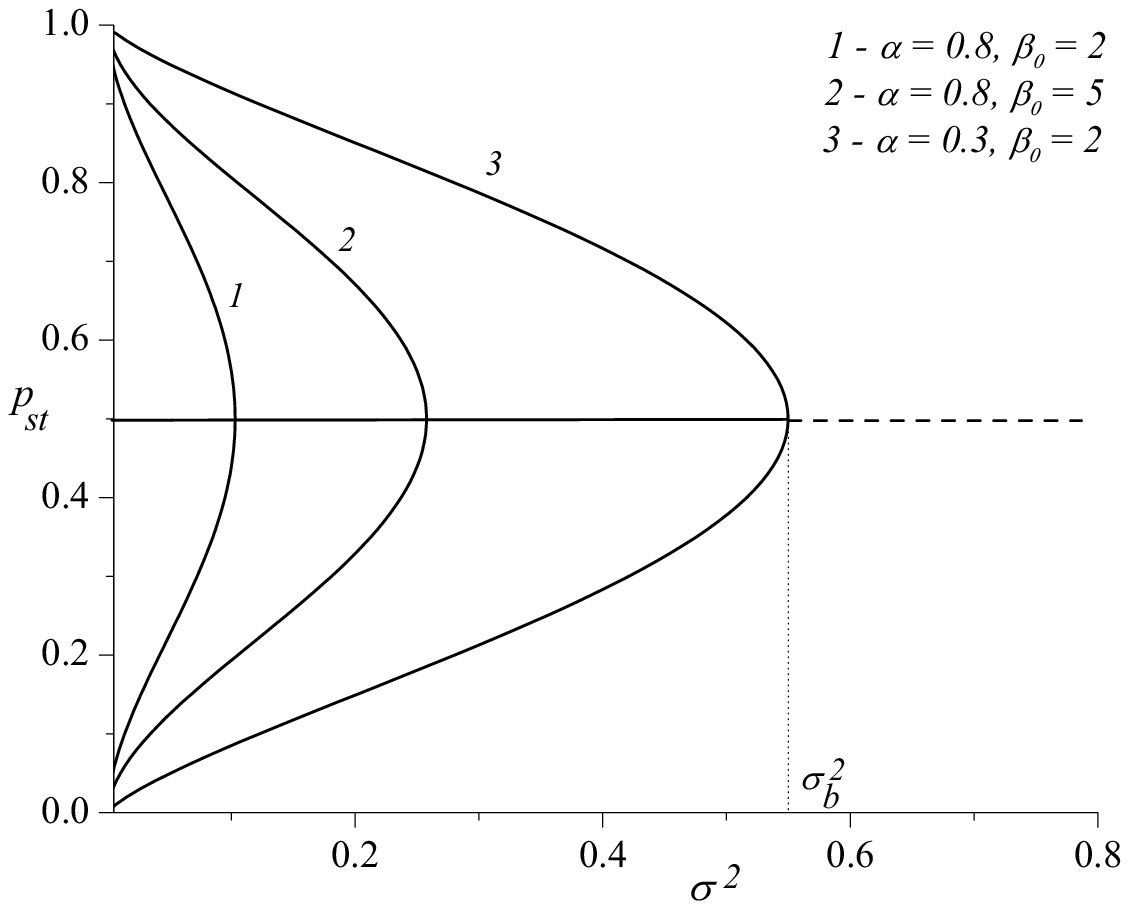}
 \small{b)}\includegraphics[width=65mm]{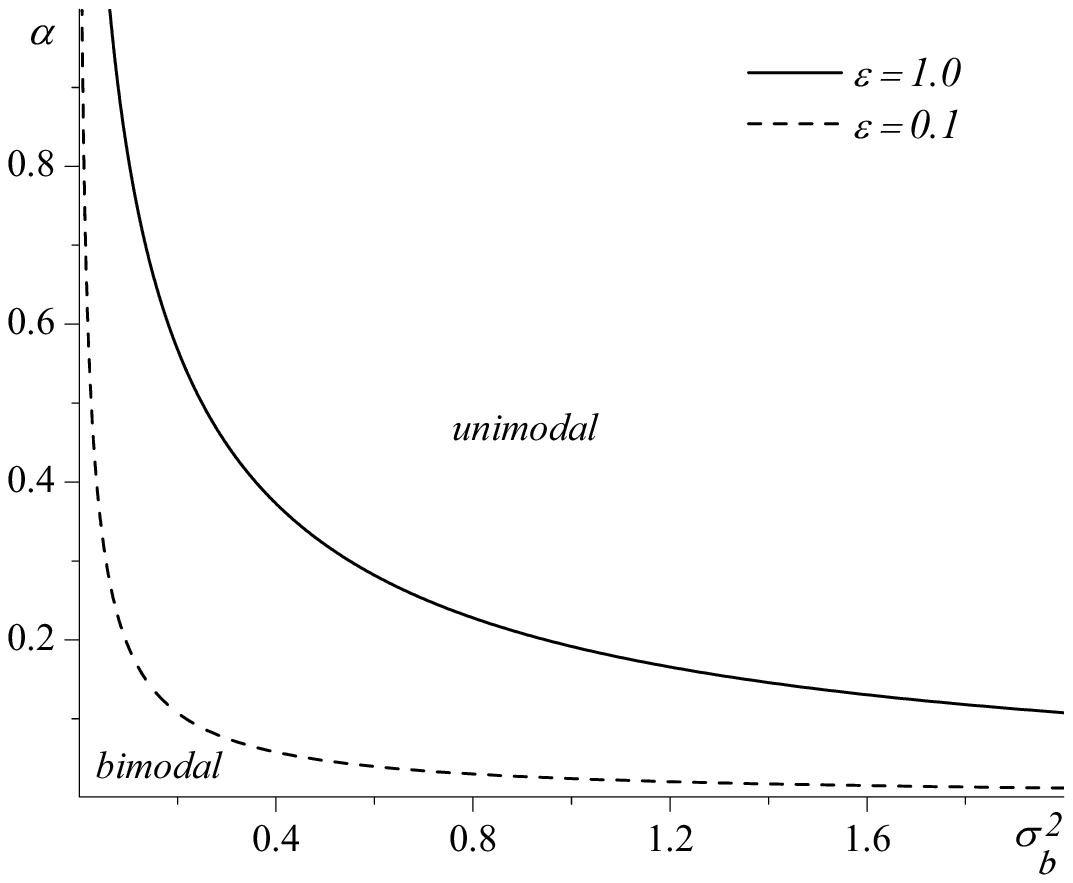}
\caption{Most probable values of the quantity $p$ versus the noise intensity
$\sigma^2$ at $\ve=1.0$ (a) and phase diagram of a system at $\beta_0=2.0$ (b)}
 \label{std_sts}
\end{figure}

Considering homogeneous states, one can put $\nabla p=\Delta p =0$ in
Eq.(\ref{Delta_p}). The reduced equation gives the most probable stationary
states $p_e$, defining extrema positions of the function $U_{ef}(p)$. The
corresponding solutions of such equation are shown in Fig.\ref{std_sts}a.
It is seen that with an increase in the noise intensity two different values of
the most probable concentration degenerate at a bifurcation point
$\sigma^2=\sigma^2_b$. Despite the fixed point $p_e=1/2$ exists always, the two
additional fixed points (upper and lower curves in Fig.\ref{std_sts}a) are
observed at $\sigma^2<\sigma^2_b$. At $\sigma^2>\sigma^2_b$ the stationary
probability function has one extremum only located in $p_{e}=1/2$. The phase
diagram $\alpha(\sigma^2_b)$ illustrating the dependence of the bifurcation
point position at various values of $\ve$ is shown in Fig.\ref{std_sts}b. One
can see that with an increase in the exponent $\alpha$ the bifurcation value
$\sigma_b^2$ decreases. Therefore, if $D_{ef}$ decreases sharply at $p\simeq0$
and $p\simeq1$, then the stationary distribution becomes unimodal at large
values of the noise intensity $\sigma^2$.

Next let us investigate stationary structures $p_{st}$. Firstly, we consider
the vicinity of the homogeneous solution $p_{e}=1/2$. In a linear approximation
one can put $\left(\nabla p\right)^2=0$ in Eq.(\ref{Delta_p}) and expand the
left part of Eq.(\ref{Delta_p}). Then, we obtain the second order differential
equation in the form
\begin{equation}
\frac{\partial^2 p}{\partial r^2}\simeq A k_l^2\left(p-p_{st}\right),\quad
 \begin{array}{c}
  A=+1\ {\rm if}\ \sigma^2>\sigma^2_b,\\
  A=-1\ {\rm if}\ \sigma^2<\sigma^2_b
 \end{array},
\end{equation}
where $k_l=\sqrt{(\ve-\tilde{\sigma}^2)/\varkappa^2}$. So, one can conclude,
that at $\sigma^2<\sigma^2_b$ the stationary state $p_e=1/2$ is locally stable
and represents a center in the phase space ($p,\nabla p$); in the opposite case
($\sigma^2>\sigma^2_b$) it changes the stability and becomes a saddle. It is
seen that the dependence $k_l(\tilde\sigma^2)$ has monotonically decreasing
character. As it follows from our consideration the stationary periodic
structures are formed in the vicinity of the point related to maximum of the
function $U_{ef}(p)$.

\begin{figure}[!ht]
\centering
\includegraphics[width=65mm]{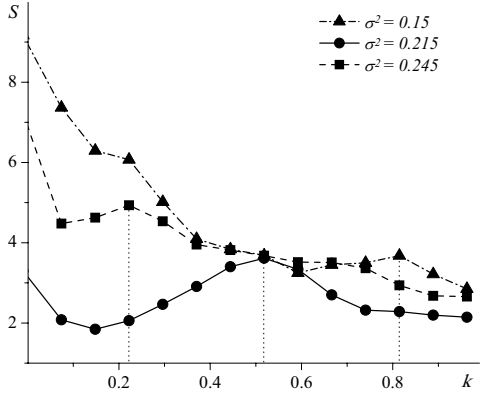}
\caption{Dependence of structure function $S$ vs. wave vector ${\bf k}$ at
$\ve=1.0$ $\alpha=0.5$, $\beta_0=2.0$ and various values of noise intensity}
 \label{s(w)}
\end{figure}

In the numerical studying of the noise-induced spatial patterns we integrate
Eq.(\ref{NFPE2}) on a $d$-dimensional lattice of the mesh size $l$. In the
discrete space Eq.(\ref{NFPE2}) can be written as follows:
\begin{equation}
 \begin{split}
 \frac{\D p_i}{\D t}=f(p_{i})&+ \frac{1}{4l^2d}\sum\limits_{j\in nn^+(i)}(p_j-p_i)^2 \frac{\D D_{ef}(p_{i})}{\D p_i}+
 D_{ef}(p_{i}) \sum\limits_{j\in nn(i)}\mathcal{D}_{ij}p_j\\
 &-\frac{\sigma^2}{2}\frac{\D D_{ef}(p_{i})}{\D p_i}\frac{1}{D^2_{ef}(p_{i})}
 +\frac{1}{\sqrt{D(p_{i})}}\xi,
 \end{split}
\label{langeven}
\end{equation}
where $i=1,\ldots,N^d$ enumerates the element of the square lattice of $N^d$
cells; periodic boundary conditions are used. The second term in
Eq.(\ref{langeven}) represents approximation of the gradient $|\nabla p |^2$,
where $nn^+(i)$ indicates nearest neighbors in the positive direction of each
axis, whereas the third term is related to a discrete Laplacian ($\nabla p \to
\sum\limits_{j\in nn(i)}\mathcal{D}_{ij}p_j$, $nn(i)$ denotes nearest neighbors
of the site $i$). To describe spatial patterns we use a spherical averaging of
the structure function $S(k,t)=\int_{\Gamma_k} S({\bf k},t)\D\Gamma$, where
$\Gamma_k$ is a spherical shell of a radius $k$. A convenient formula is
\begin{equation}
S(k,t)=\frac{1}{N_k}\sum\limits_{k\leq k\leq k+\Delta k}S({\bf k},t).
\end{equation}
All calculations were performed in a two-dimension square lattice of $120\times
120$ cells with lattice scaling $l=1$, and integration time step $\tau=5\times
10^{-3}$. The stationary spherically averaged structure function is shown in
Fig.\ref{s(w)} at different values of the noise intensity.

It is principle, that an increase in the noise intensity $\sigma^2$ results to
a shift of the peak of $S(k)$ to the small values of the wave vector ${\bf k}$.
This is different from what happens at the early stages of evolution. This fact
indicates that when the system starts to evolve a multiplicative noise leads to
instability of the state $p=1/2$; the corresponding values of the wave vector
${\bf k}$ increase with the noise intensity growth. For large times (in the
stationary case) the situation is quite different. Here with an increase in the
noise intensity the peak of $S(k)$ moves toward smaller values of ${\bf k}$.
This fact is related to an entropy mechanism of the phase transitions. Indeed,
the effective energy functional (\ref{U_eff}) is defined through the initially
known free energy functional $F[p]$ and a contribution $-\sigma^2/2\int\D {\bf
r}\ln D_{ef}(p)$. The last one gives an effective entropy $S_{ef}=-1/2\int\D
{\bf r}\ln D_{ef}(p)$ multiplied by the noise intensity $\sigma^2$. In such a
case we arrive at the standard thermodynamic definition of an internal energy
$U_{ef}=F+\sigma^2S_{ef}$. It is well known that in the theory of the
entropy-driven phase transitions self-organization processes are not related to
the short-time instability of the mixed/disordered state \cite{GO2001}, its
caused by entropy variations following from concentration-dependent mobility.
In our case we have a quite similar situation. In particular, at early stages
of the system evolution a noise destabilizes the disordered state $p=1/2$,
whereas at $t\to\infty$ the entropy contribution plays a crucial role: it leads
to patterning with small values of ${\bf k}$. Obtained numerical results are in
good corresponding with analytical predictions: in the stationary case the peak
of $S(k)$ moves toward small values of the wave vector when the noise intensity
increases.

\section{Mean Field Theory}

\begin{figure}[!ht]
\centering
\includegraphics[width=65mm]{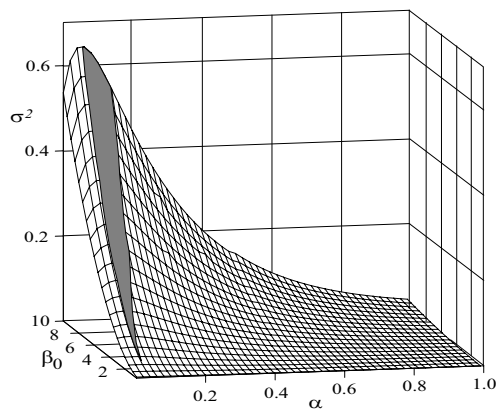}
\caption{Phase diagram at $\ve=0.1$ (shaded domain corresponds to the
non-critical phase transitions)}
 \label{phs2d}
\end{figure}

In this section we will consider the noise-induced phase transitions with the
above entropy mechanism. To this end we construct the Weiss' mean-field (MF)
approximation, based on the stationary distribution function. Let us
approximate the gradient term in the functional (\ref{U_eff}) by the sum over
nearest neighbors on the lattice as Eq.(\ref{langeven}) shows. In the MF
approximation we replace the exact value of the neighbors by a mean field
$\eta=\langle p\rangle$. It leads to the relation $\left(\nabla
p\right)^2\to\left(p-\langle p\rangle\right)^2=\left(p-\eta\right)^2$. In such
a case the quantity $\eta$ can be used as an order parameter. In this procedure
we neglect fluctuations in neighboring sites. The value of the order parameter
can be computed self-consistently as follows:
\begin{equation}
\eta\equiv\langle p\rangle=\int P_{st}(p;\eta)p\D p \equiv\Phi(\eta),
\end{equation}
where $P_{st}(p;\eta)$ is the stationary distribution in the MF approximation
depending on the mean field $\eta$. As usual, the parameter $\eta$ allows to
identify a phase transition from a disordered state with $\eta=\eta_c$ to an
ordered one characterized by the value $\eta\ne\eta_c$; the symmetry that leads
to $\eta=\eta_c$ is embedded in the system. According to the discussion above,
the effective potential has the form
\begin{equation}
U_{ef}(p;\eta)=\frac{1}{2}D_{ef}^2(p)(p-\eta)^2-\frac{\sigma^2}{2}\ln
D_{ef}(p)-\int\limits_0^pf(p')D_{ef}(p')\D p' \label{potencial}
\end{equation}
and depends on the mean field $\eta$. The stationary probability density
function is
\begin{equation}
P_{st}(p;\eta)=\mathcal{Z}(\eta)\exp\left(-U_{ef}(p;\eta)/\sigma^2\right),
\end{equation}
where $\mathcal{Z}(\eta)$ takes care of the normalization condition $\int
P_{st}(p;\eta)\D p=1$. According to formalism proposed in Ref.\cite{katya} and
quantitative analysis of the function $\Phi(\eta)$ one can expect the
second-order phase transitions in a system under consideration. Using
Newton-Raphson condition, transitions between disordered and ordered phases
occur if
\begin{equation}
\left.\frac{\D\Phi(\eta)}{\D\eta}\right|_{\eta=\eta_c}=1,\quad \eta_c=1/2.
 \label{cond}
\end{equation}
Introducing the susceptibility in the form
\begin{equation}
\chi=\frac{\D\Phi(\eta)}{\D\eta}\label{chi},
\end{equation}
we relate the critical points to the values of $\sigma^2$ when $\chi=1$. The
corresponding phase diagram is shown in Fig.\ref{phs2d}.

\begin{figure}[!ht]
\centering
 \small{a)}\includegraphics[width=65mm]{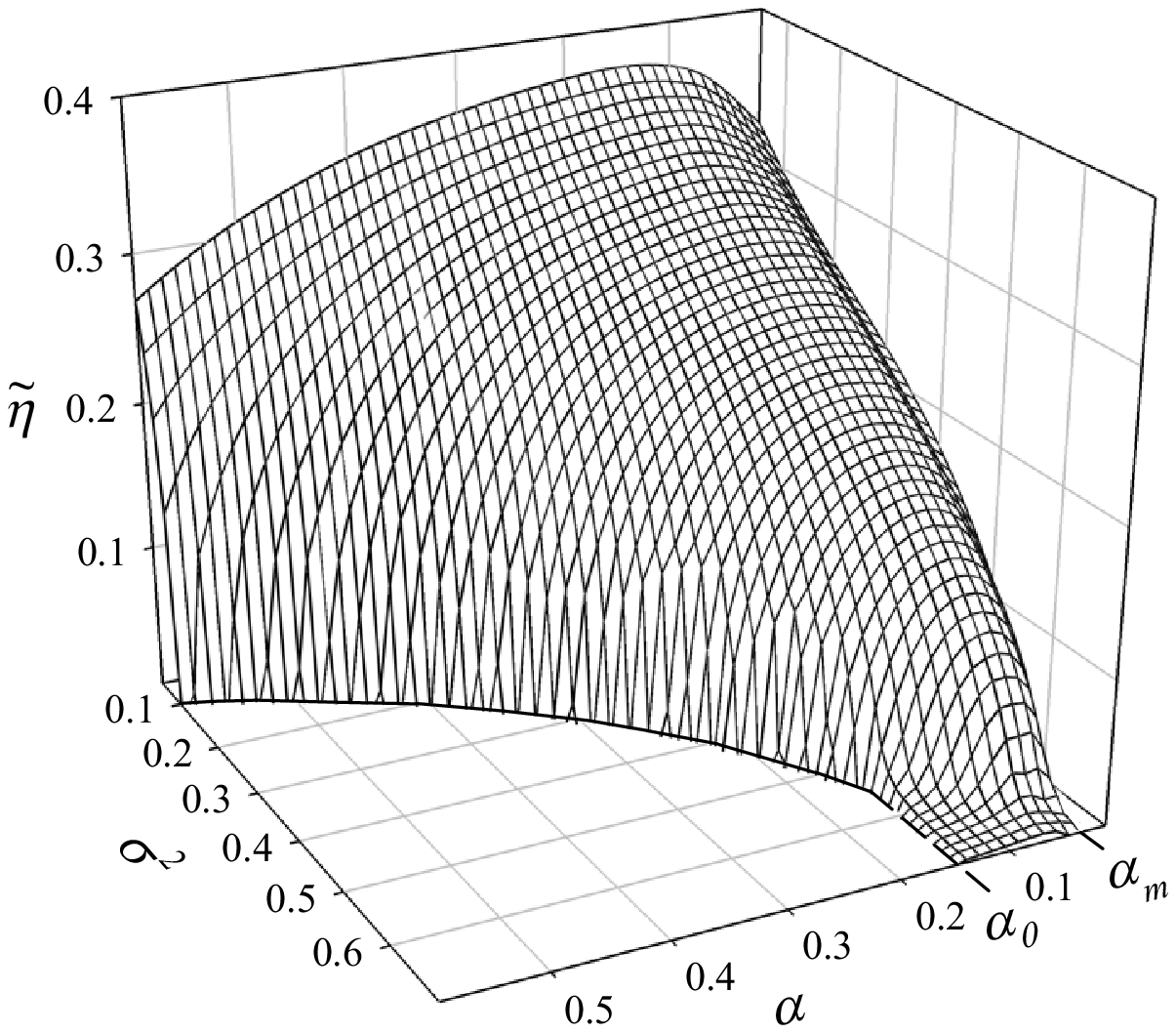}
 \small{b)}\includegraphics[width=65mm]{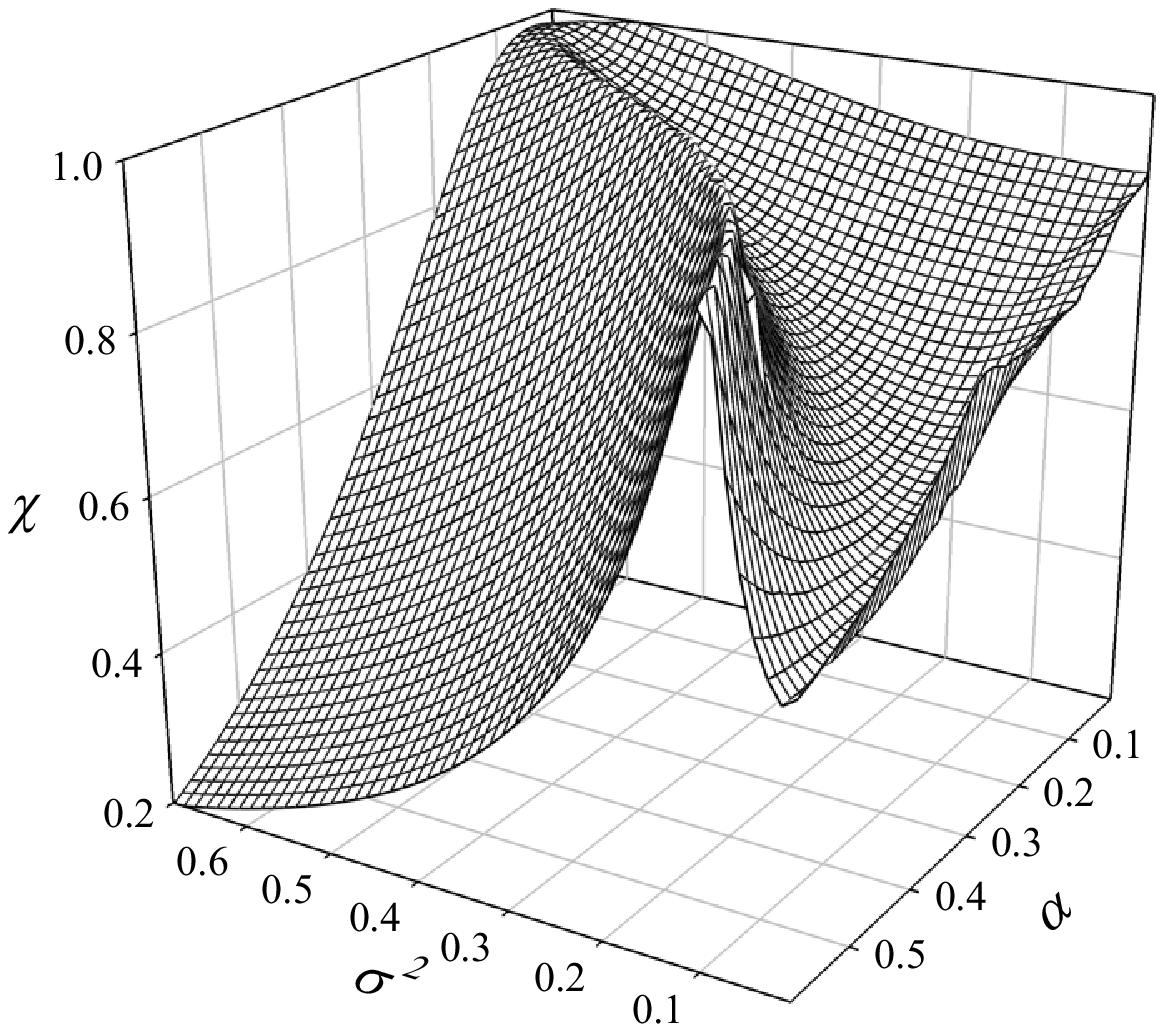}
\caption{The character of changing of the order parameter and the
susceptibility at $\ve=0.1$, $\beta_0=10.0$} \label{grf_eta}
\end{figure}

Here the surface plotting with the help of the solid lines corresponds to the
critical values of the system parameters, when the second-order phase
transition are occur. Shaded domain corresponds to the phase transitions which
are of non-critical character (such situation is observed in magnetic systems
when a weak external field is introduced into the system). Here the order
parameter decreases continuously and has not singularity in the vicinity of the
transition point related to the shaded domain. The related susceptibility has
the broad peak in the point corresponding to the system parameters values of
the shaded domain shown in Fig.\ref{phs2d}. To understand a principle change in
the system behaviour one can estimate contributions of all nonlinearities
appeared in the effective potential (\ref{potencial}). Here one has a
competition between the noise characterized by $D_{ef}(p)$ and the driving
force $f(p)$. Indeed, at $\alpha\simeq 1$ as the term related to the force
$f(p)$ as the term related to $\ln D_{ef}(p)$ are essential and lead to phase
transition of a critical character. At $\alpha\ll 1$ the main nonlinearity is
caused by the force $f(p)$ only. Here the system can be approximately described
by $D_{ef}(p)\approx {\rm const}$. At the $\alpha$-values related to the shaded
domain there are two above nonlinear terms having weak nonlinearities than the
term related to $D_{ef}^2(p-\eta)^2$ gives. Therefore, at such intermediate
values of $\alpha$ there are no nonlinearities suppressing the order in the
system. A size of the domain for such transitions
$\Delta\alpha=\alpha_0-\alpha_m$ depends on control parameters $\beta_0$ and
$\ve$. Let us introduce an effective order parameter
$\tilde{\eta}=\left|1/2-\eta\right|$: in the ``disordered'' phase with $p=1/2$
one has $\tilde\eta=0$; in the ``ordered'' phase $\tilde\eta\ne0$. The
dependencies of the effective order parameter and the susceptibility versus the
exponent $\alpha$ of the effective diffusion coefficient $D_{ef}$ and noise
intensity $\sigma^2$ are shown in Fig.\ref{grf_eta}.
From the Fig.\ref{grf_eta}a one can see that at $\alpha>\alpha_0$ and
$\alpha<\alpha_m$ the effective order parameter $\tilde\eta$ decreases
critically to zero with the noise intensity growth. In the domain
$\alpha\in(\alpha_m,\alpha_0)$ the dependence $\tilde\eta(\sigma^2)$ has a
smooth falling-down character. Therefore, such transitions are non-critical.
Figure \ref{grf_eta}b illustrates peak of the susceptibility $\chi$ that
corresponds to the value $\chi(\alpha,\sigma^2)=1$ and determines the
second-order phase transition point. In the interval of the exponent
$\alpha\in(\alpha_m,\alpha_0)$ the susceptibility $\chi$ has the broad peak
which is characterizes by value $\chi\simeq 1$ and relates to non-critical
phase transitions. Thus, as it follows from our calculations to obtain the
critical (second-order) noise-induced phase transitions the exponent $\alpha$
should lie in the interval $(0,\alpha_m)\bigcup(\alpha_0,1)$.

\begin{figure}[!ht]
\centering
 \small{a)}\includegraphics[width=65mm]{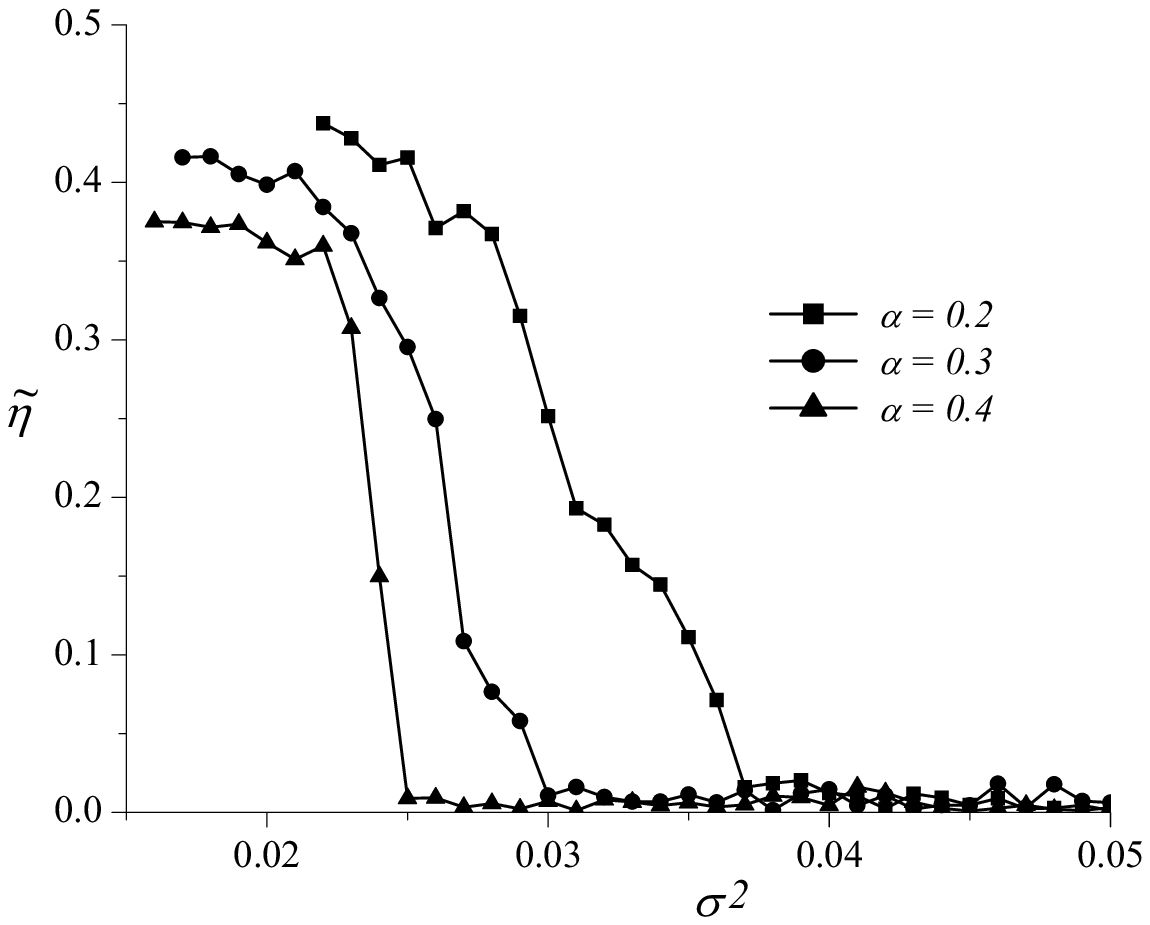}
 \small{b)}\includegraphics[width=65mm]{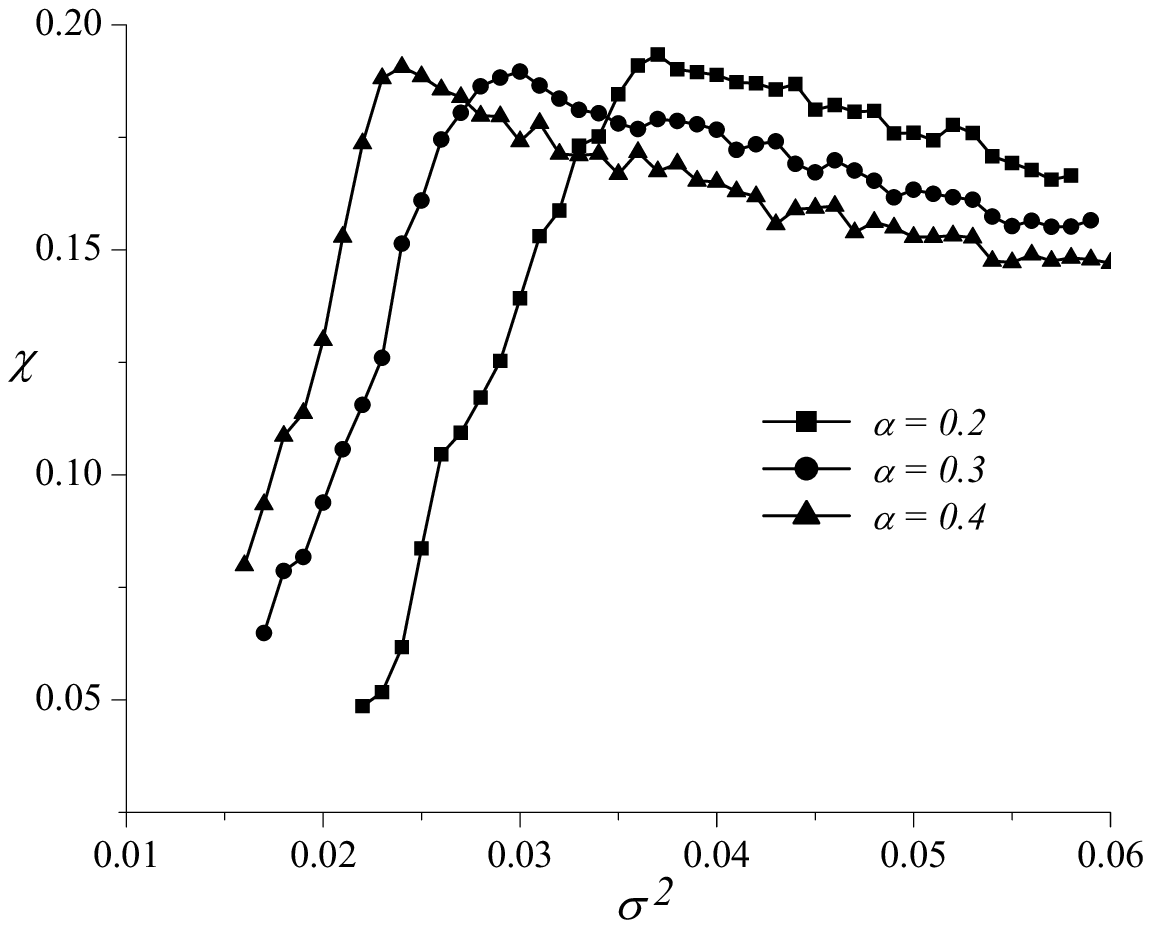}
\caption{The dependence of the order parameter $\tilde{\eta}$ and the
susceptibility $\chi$ vs. noise intensity $\sigma^2$ at $\ve=0.1$,
$\beta_0=3.0$ } \label{exp_eta}
\end{figure}

To verify our MF results we integrate numerically Langevin equation
(\ref{NFPE2}) on the two-dimensional lattice according to algorithm presented
above (see Eq.(\ref{langeven})). For the order parameter we use the formula
$\tilde\eta=\left|1/2-\langle p\rangle\right|$; the generalized susceptibility
measures fluctuations of the field $p$ in the vicinity of the critical points,
$\chi\simeq\langle\Delta p\rangle^2=\langle p^2 \rangle-\langle p \rangle^2$,
averaging over noise realizations, time, and the ensemble is taking into
account. Solving the discrete Langevin equation Eq.(\ref{langeven}) numerically
on a two-dimension square lattice of $120\times 120$ cells with mesh size
$l=1$, and integration time step $\tau=5\times 10^{-3}$, we obtain dependencies
$\tilde{\eta}(\sigma^2)$ and $\chi(\sigma^2)$ shown in Fig.\ref{exp_eta}.
It is seen, that the order parameter decreases critically with the noise
intensity growth. The generalized susceptibility has a well pronounced peak
located at the critical transition point. With an increase in the exponent
$\alpha$ the peak location of the generalized susceptibility $\chi$ is shifted
toward small values of the noise intensity $\sigma^2$. Our numerical results
are in a good correspondence with the MF theory.

\begin{figure}[!ht]
\centering
\includegraphics[width=70mm]{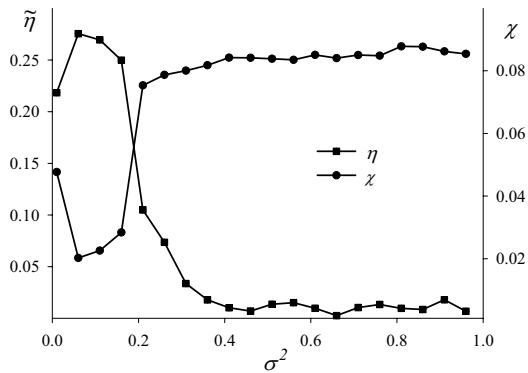}
\caption{The dependence of the order parameter $\tilde{\eta}$ and the
susceptibility $\chi$ vs. noise intensity $\sigma^2$ at $\ve=0.1$,
$\alpha=0.1$, $\beta_0=10.0$ } \label{exp_non}
\end{figure}

Figure \ref{exp_non} illustrates dependence of the order parameter and the
generalized susceptibility at the system parameters values related to the
shaded domain in Fig.\ref{phs2d}.
It is seen that the order parameter decreases smoothly than in
Fig.\ref{exp_eta}a; the generalized susceptibility has no pronounced peak.
Thus, one can conclude that the corresponding phase transition is of the
non-critical character.

\section{Conclusions}

In this paper we have considered the physical system of the reaction-diffusion
kind with a field-dependent mobility and internal fluctuations. It was shown
that in generalized approach basing on the nonlinear kinetic equation the
field-dependent mobility is of a power-low form proposed by C.L. Emmott and
A.J. Bray \cite{Bray}. Considering the local dynamics of the concentration
field, it was found that internal fluctuations can sustain stable periodic
patterns if the noise intensity does not exceed a critical value. The
mechanisms of the phase transitions have been studied with the help of the mean
field theory. It was shown that an increase in the noise intensity leads to
disordering phase transitions. We have found that there is a special domain of
the exponent determining the power-low form of the field-dependent diffusion
coefficient when disordering phase transitions are not critical. It can be
explained by the competition between noise and driving force. The corresponding
bifurcation and phase diagrams are verified by computer simulations.

\section*{Acknowledgement}

The author is thankful to Profs. E.D. Belokolos and D.O. Kharchenko for
inspiring discussions and helpful comments, and to Institute of Applying
Physics, Nat. Acad. Sci. of Ukraine for the vested possibility of using the
computer cluster to perform computer simulations.

\end{document}